\documentclass[intlimits,twoside,a4paper]{article}
\usepackage{wrapfig}
\usepackage{graphicx}
\usepackage[T2A]{fontenc}
\usepackage{amssymb}
\usepackage[cp1251]{inputenc}

\usepackage[]{cmpj2}
\articletype{Rapid communication}

\newcommand{\be}{\begin{equation}}
\newcommand{\ee}{\end{equation}}
\newcommand{\bea}{\begin{eqnarray}}
\newcommand{\eea}{\end{eqnarray}}
\newcommand{\non}{\nonumber\\}

\newcommand{\dtwo}[2]{\frac{\partial^2 #1}{\partial #2^2}}
\newcommand{\done}[2]{\frac{\partial #1}{\partial #2}}
\newcommand{\eps}{\varepsilon}

\title[Piezoelectric Resonance in KH$_2$PO$_4$ Type
Crystals Revisited]{Piezoelectric Resonance in KH$_2$PO$_4$ Type
Crystals Revisited}
\author[R.R. Levitskii,
    I.R. Zachek, A.P. Moina, A.S. Vdovych]{R.R. Levitskii\refaddr{label1},
    I.R. Zachek\refaddr{label2}, A.P. Moina\refaddr{label1}, A.S. Vdovych\refaddr{label1}}
\addresses{
\addr{label1} Institute for Condensed Matter Physics of the
National Academy of Sciences of Ukraine, 1 Svientsitskii str,
79011 Lviv, Ukraine \addr{label2} Lviv Polytechnic National
University, 12 Bandera Str., 79013 Lviv, Ukraine }
\begin{document}

\maketitle

\begin{abstract}
Within the framework of proton model with taking into account the
piezoelectric interaction with the shear strain $\varepsilon_6$, a
dynamic dielectric response of KH$_2$PO$_4$ family ferroelectrics
and antiferroelectrics is considered. Piezoelectric resonance
frequencies of rectangular thin plates of the crystals cut in the
(001) plane (0$^\circ$ Z-cut) are calculated, which are found to
be in a good agreement with experiment.
\pacs 77.22.Ch, 77.22.Gm,  77.84.Fa, 77.65.Fs

\end{abstract}

\section{Introduction}

In our previous papers \cite{old,old-adp} we explored the dynamic
dielectric response of square thin plates cut from the
KH$_2$PO$_4$ family crystals in the planes (001), perpendicular to
the axis of spontaneous polarization. Using the modification of
the proton ordering model \cite{our!!} that includes the
piezoelectric coupling with the shear strain $\eps_6$, within the
framework of the Glauber approach \cite{61x} and the four-particle
cluster approximation, we obtained expressions for the dynamic
dielectric permittivity of the crystals, which took into account
the dynamics of the shear strain $\eps_6$. In the low-frequency
limit these expressions coincided with the static permittivities
of mechanically free crystals, whereas in the microwave region
they coincided with the dynamic permittivities of clamped
crystals, exhibiting a relaxational dispersion.

In the intermediate region, the obtained permittivities had
numerous peaks associated with the piezoelectric resonances.
However, while solving the partial differential equations for the
strain in \cite{old,old-adp}, the boundary conditions were not set
correctly. Instead of demanding that the entire edges of the plate
were mechanically free, we considered the plate free only at its
vertices. It resulted in the underestimated values of the resonant
frequencies. In the present paper we shall correct these errors.

We shall not repeat here the details of the previous calculations,
which were correct. The system Hamiltonians, most of the used
notations, as well as derivation of the dynamic dielectric
permittivities of clamped crystal (the pseudospin subsystem
dynamics), can be found in \cite{old,old-adp}.

\section{Dynamic permittivity of KH$_{2}$PO$_{4}$ type crystals}

We shall consider shear mode vibrations of a thin $L_x \times L_y$
rectangular plate of a KH$_2$PO$_4$ crystal, cut in the (001)
plane, with the edges along [100] and [010] ($0^\circ$ Z-cut). The
vibrations are induced by time-dependent electric field
$E_{3t}=E_3e^{i\omega t}$. In the ferroelectric phase this field,
in addition to the shear strain $\varepsilon_6$, induces also the
diagonal components of the strain tensor $\varepsilon_i$, but for
the sake of simplicity we shall neglect them.

Dynamics of pseudospin subsystem will be considered in the spirit
of the stochastic Glauber model \cite{61x}, using the
four-particle cluster approximation.  The  system of equations for
the time-dependent deuteron (pseudospin) distribution functions is
  \be
  \label{3.1}
  - \alpha \frac{d}{dt} \langle \prod_f \sigma_{qf} \rangle =
  \sum\limits_{f'} \langle \prod_f \sigma_{qf} \left[ 1 -
  \sigma_{qf'} \tanh \frac12 \beta {\mbox{\boldmath$\varepsilon$}}_{qf'}^z(t)
  \right] \rangle ,
  \ee
where ${\mbox{\boldmath$\varepsilon$}}_{qf'}^z(t)$ is the local
field acting on the  $f'$th
 deuteron in the $q$th cell, which can be found from the system Hamiltonian (see
 \cite{old}); $\alpha$ is the parameter setting the time scale of
 the dynamic processes in the pseudospin subsystem.

Taking into account the symmetry of the distribution functions
 \bea
 && \hspace{-4ex} \eta^{(1)} = \langle \sigma_{q1} \rangle =  \langle \sigma_{q2}
 \rangle =  \langle \sigma_{q3} \rangle =  \langle \sigma_{q4}
 \rangle, \non
 && \hspace{-4ex} \eta^{(3)} = \langle \sigma_{q1}  \sigma_{q2} \sigma_{q3} \rangle =  \langle \sigma_{q1}
 \sigma_{q3} \sigma_{q4}  \rangle =  \langle \sigma_{q1}  \sigma_{q2} \sigma_{q4}
 \rangle = \langle \sigma_{q2}  \sigma_{q3} \sigma_{q4} \rangle, \\
 && \hspace{-4ex} \eta_1^{(2)} \!=\! \langle \sigma_{q2} \sigma_{q3} \rangle \!=\! \langle \sigma_{q1} \sigma_{q4}
 \rangle, ~~~~  \eta_2^{(2)} \!=\! \langle \sigma_{q1} \sigma_{q2} \rangle \!=\! \langle \sigma_{q3} \sigma_{q4}
 \rangle, ~~~~  \eta_3^{(2)} \!=\! \langle \sigma_{q1} \sigma_{q3} \rangle \!=\! \langle \sigma_{q2} \sigma_{q4}
 \rangle,
 \nonumber
 \eea
from (\ref{3.1}) we obtain for them a closed system of equations
 \be
 \label{3.5}
 \alpha \frac{d}{dt} \left(\!\! \begin{array}{ccccc}
                        \eta^{(1)}\\
                        \eta^{(3)}\\
                        \eta_1^{(2)}\\
                        \eta_2^{(2)}\\
                        \eta_3^{(2)}
                        \end{array}
                        \!\!\right) \!=\!  \left(\!\! \begin{array}{ccccc}
                                \bar c_{11} & \bar c_{12} & \bar c_{13} & \bar c_{14} & \bar c_{15}  \\
                                \bar c_{21} & \bar c_{22} & \bar c_{23} & \bar c_{24} & \bar c_{25}  \\
                                \bar c_{31} & \bar c_{32} & \bar c_{33} & \bar c_{34} & \bar c_{35}  \\
                                \bar c_{41} & \bar c_{42} & \bar c_{43} & \bar c_{44} & \bar c_{45}  \\
                                \bar c_{51} & \bar c_{52} & \bar c_{53} & \bar c_{54} & \bar c_{55}
                                \end{array}
                                \!\!\right)
                        \left(\!\! \begin{array}{ccccc}
                        \eta^{(1)}\\
                        \eta^{(3)}\\
                        \eta_1^{(2)}\\
                        \eta_2^{(2)}\\
                        \eta_3^{(2)}
                        \end{array}
                        \!\!\right) \!+\!
                                \left(\!\! \begin{array}{ccccc}
                                \bar c_1 \\
                                \bar c_2 \\
                                \bar c_3 \\
                                \bar c_4 \\
                                \bar c_5
                                \end{array}
                                \!\!\right).
 \ee
The used here notations can be found in \cite{old}.

Dynamics of the deformational processes is described using
classical Newtonian equations of motion of an elementary volume,
which for the relevant to our system displacements $u_1$ and $u_2$
($\varepsilon_6 =  \frac{\partial u_1}{\partial
 y} + \frac{\partial u_2}{\partial x}$) read
\begin{equation}
\label{3.6} \rho\dtwo{u_1}{t}=\done{\sigma_{6}}{y},\quad
\rho\dtwo{u_2}t=\done{\sigma_{6}}{z}.
\end{equation}
Here $\rho$ is the crystal density,  $\sigma_{6}$ is the
mechanical shear stress,  which, being the function of
$\eta^{(1)}$, $E_3$, and $\varepsilon_6$, is found from the
constitutive equations derived in
 \cite{old}.

 At small deviations from the equilibrium we can separate in the systems
 (\ref{3.5}) and (\ref{3.6}) the static and time-dependent parts, presenting the dynamic variables $\eta^{(1)}$, $\eta^{(3)}$,
 $\eta^{(2)}_i$, $ \varepsilon_6$,
 $u_{1,2}$ as sums of the equilibrium
values and of their fluctuational deviations, while the
fluctuational parts are assumed to be in the form of harmonic
waves
 \begin{eqnarray}
 && \eta^{(1)} = \tilde \eta^{(1)} +  \eta^{(1)}(x,y)e^{i\omega t}, ~~
 \eta^{(3)} = \tilde \eta^{(3)} +\eta^{(3)}(x,y)e^{i\omega
 t}\ldots
 \nonumber
 \end{eqnarray}
The fluctuational part of  (\ref{3.5}) is then reduced to the
system of linear first-order differential equations with constant
coefficients, solving which we get
 \begin{eqnarray}
 \label{eta1}
 && \eta^{(1)}(x,y) = \frac{\beta \mu_3}{2} F^{(1)}(\alpha
 \omega)E_3 + \Bigl[ - \beta \psi_6F^{(1)}(\alpha \omega) + \\
 && + \beta \delta_{s6} F_s^{(1)}(\alpha \omega) - \beta \delta_{a6} F_a^{(1)}(\alpha \omega) + \beta \delta_{16} F_1^{(1)}(\alpha
 \omega)\Bigr] \varepsilon_{6}(x,y), \nonumber
 \end{eqnarray}
the notations introduced here can be found in \cite{old}.

Substituting (\ref{eta1}) into Eqs. (\ref{3.6}), we obtain
 \begin{eqnarray}
 \label{3.7}
 && \frac{\partial^2 u_{1}}{\partial y^2} + k_{6}^2u_{1} = 0,
 ~~~ \frac{\partial^2 u_{2}}{\partial x^2} + k_{6}^2u_{2} = 0,
 \end{eqnarray}
where $k_{6}$ is the wavevector
 \be
 \label{7.14}
 k_{6} = \frac{\omega
 \sqrt{\rho}}{ \sqrt{c_{66}^E(\alpha\omega)} },
  \ee
  and
   \begin{eqnarray}
 &&  c_{66}^E(\alpha \omega) = c_{66}^{E0} + \frac{4\beta \psi_6}{v D_6}f_6 + \frac{2\beta}{v D_6^2} (- \delta_{s6}M_{s6} + \delta_{16}M_{16} +
 \delta_{a6}M_{a6})^2 + \label{c66} \\
 && + \frac{4\beta
 \psi_6}{v} \Bigl[ - \psi_6 F^{(1)}(\alpha \omega) +
 \delta_{s6}F_s^{(1)}(\alpha \omega) + \delta_{16}F_1^{(1)}(\alpha
 \omega) - \delta_{a6}F_a^{(1)}(\alpha \omega) \Bigr] - \non
 &&  - \frac{4\varphi_3 f_6}{v D_6} \beta \Bigl[ - \psi_6F^{(1)}(\alpha
 \omega) + \delta_{s6}F_s^{(1)}(\alpha \omega) + \delta_{16}F_1^{(1)}(\alpha
 \omega) - \delta_{a6}F_a^{(1)}(\alpha \omega) \Bigr] - \non
 &&  - \frac{2\beta}{v D_6}
 \Bigl[ \delta_{s6}^2 \cosh (2\tilde z + \beta \delta_{s6} \tilde
 \varepsilon_6)+ 4b\delta_{16}^2\cosh (\tilde z - \beta \delta_{16} \tilde
 \varepsilon_6) + \delta_{a6}^22a\cosh \beta \delta_{a6} \tilde
 \varepsilon_6^2 \Bigr]. \nonumber
 \end{eqnarray}

  Differentiating the first and second equations of (\ref{3.7}) with
  respect to $y$ and $x$, correspondingly, remembering that we
  neglect the diagonal strains $\varepsilon_1=\partial
  u_1/\partial x$ and $\varepsilon_2=\partial
  u_1/\partial y$, and adding the two obtained equations, we
  arrive at the single equation for the strain $\eps_6$
\begin{eqnarray}
&& \label{singlesystem}
\dtwo{\varepsilon_{6}(x,y)}{x}+\dtwo{\varepsilon_{6}(x,y)}{y}+k_6^2\varepsilon_{6}(x,y)=0.
\end{eqnarray}
Boundary conditions for $\varepsilon_{6}(x,y)$ follow from the
assumption that the crystal is simply supported, that is, it is
traction free at its edges (at $x=0$, $x=L_x$, $y=0$, $y=L_y$, to
be denoted as $\Sigma$)
\begin{equation}
\label{boundary1} \sigma_6|_\Sigma=0.
\end{equation}
In our previous consideration \cite{old} this condition was
fulfilled at the corners of the crystal plate only, but not along
all its edges.  Substituting (\ref{boundary1}) into the
constitutive relations, we obtain the explicit boundary conditions
for the strains in the following form
\begin{equation}\label{boundary2}
\varepsilon_{6}|_\Sigma\equiv\varepsilon_{i0}=
\frac{e_{36}(\alpha\omega)}{c_{66}^E(\alpha\omega)}E_3,
\end{equation}
 where
 \begin{eqnarray}
 \label{e36}
&& e_{36} (\alpha\omega) = e_{36}^0 +
 \frac{\beta\mu_3}{v} \Bigl[ -\psi_6 F^{(1)}(\alpha\omega) +
 \delta_{s6}F_s^{(1)}(\alpha\omega) +  \delta_{16}F_1^{(1)}(\alpha\omega) -
 \delta_{a6}F_a^{(1)}(\alpha\omega)\Bigr].
 \end{eqnarray}
Solution of (\ref{7.14}) with the boundary conditions
(\ref{boundary2}) is
\begin{equation}
\label{strain}
\eps_{6}(x,y)=\eps_{60}
+ \eps_{60}\sum_{k,l=0}^\infty
\frac{16}{(2k+1)(2l+1)\pi^2}\frac{\omega^2}{(\omega_{kl}^0)^2-\omega^2}\sin\frac{\pi(2k+1)x}{L_z}\sin\frac{\pi(2l+1)y}{L_y},
\end{equation}
with $\omega^0_{kl}$ given by
\begin{equation}
\label{res4} \omega_{kl}^0=\sqrt{\frac{
c_{66}^E(\omega^0_{kl})\pi^2}{\rho}\left[\frac{(2k+1)^2}{L_x^2}+\frac{(2l+1)^2}{L_y^2}\right]}.
\end{equation}

Using the expression, relating polarization $P_3$ to the order
parameter $\eta^{(1)}$ and strain $\varepsilon_6$ (see
\cite{old}), we find that
 \be
 P_3(x,y,t) = P_{3}(x,y) e^{i\omega t},\quad
 P_{3}(x,y) = e_{36}(\alpha\omega)
 \varepsilon_{6}(x,y)+ \chi_{33}^{\varepsilon}(\alpha\omega)E_3,
 \ee
where
 \begin{equation}
 \label{chie}
 \chi_{33}^\eps(\alpha\omega) = \chi_{33}^{\varepsilon 0} +
 \frac{\beta\mu_3^2}{2v}F^{(1)}(\alpha\omega).
 \end{equation}
is the dynamic dielectric susceptibility of a clamped crystal.

 Now
we can calculate the dynamic dielectric susceptibility of a free
crystal
 $\chi_{33}^{\sigma}(\alpha\omega)$ as
 \begin{eqnarray}
 && \chi_{33}^{\sigma}(\omega) = \frac{1}{L_xL_y}
 \frac{\partial}{\partial E_3} \int\limits_0^{L_x}dx\int\limits_0^{L_y}
dy P_{3}(x,y),
 \end{eqnarray}
obtaining
 \begin{equation}
 \label{dyn_susc}
\chi_{33}^\sigma(\omega)=\chi_{33}^{\eps}(\alpha\omega) +
R_6(\omega)\frac{e_{36}^2(\alpha\omega)}{c_{66}^E(\alpha\omega)},
\end{equation}
 where
 \begin{equation}
 \label{r6}
 {R_6(\omega)}=1+ \sum_{k,l=0}^\infty
\frac{64}{(2k+1)^2(2l+1)^2\pi^4}\frac{\omega^2}{(\omega_{kl}^0)^2-\omega^2}.
 \end{equation}

In the static and the high frequency limits  from (\ref{dyn_susc})
we obtain the static susceptibility of a free crystal \cite{our!!}
and the dynamic susceptibility of a mechanically clamped crystal,
exhibiting relaxational dispersion in the microwave region. Thus,
eq. (\ref{dyn_susc}) explicitly describes the effect of crystal
clamping by high-frequency electric field.

 In the
intermediate frequency region, the susceptibility has a resonance
dispersion with numerous peaks ar frequencies where ${\rm
Re}[R_6(\omega)]\to\infty$. Frequency variation of
$c_{66}^E(\alpha\omega)$ is perceptible only in the  region of the
microwave dispersion of the dielectric susceptibility. Below this
region it is practically frequency independent and coincides with
the static elastic constant $c_{66}^E$. Since the resonance
frequencies are expected to be in the $10^4-10^7$~Hz range,
depending on temperature and sample dimensions,  the equation for
the resonance frequencies (\ref{res4}) is reduced to an explicit
expression by putting in it $c_{66}^E(\alpha\omega)\to c_{66}^E$.

Comparing (\ref{res4}) to the expression obtained previously
\cite{old} for a square $L\times L$ plate cut in the (001) plane
\[
\omega_{k}=\frac{\pi(2k+1)}{L}\sqrt{\frac{c_{66}^E(\omega_k)}{\rho}},
\]

\begin{wrapfigure}{i}{0.5\textwidth}
\centerline{\includegraphics[width=0.35\textwidth]{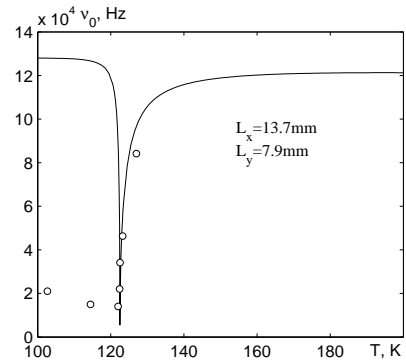}}
\caption{The first resonance frequency $\nu_0=\omega_{00}^0/2\pi$
of a rectangular 0$^\circ$ Z-cut of a KH$_2$PO$_4$ crystal.
Symbols are experimental points taken from \cite{kdp-res}. Line:
the present theory.} \label{myfig1}
\end{wrapfigure}

\noindent we can see that the incorrectly set boundary conditions
\cite{old} led to the first resonance frequency, being $\sqrt2$
times smaller than the one given by (\ref{res4}). However, the low
and high frequency limits of the susceptibility calculated in
\cite{old} (the static value and the clamped values with the
relaxational dispersion in the microwave region) were correct.

The used values of the model parameters can be found in
\cite{old}. As one can see, in the paraelectric phase the  first
resonance frequency of a rectangular 0$^\circ$ Z-cut of a
KH$_2$PO$_4$ crystal, calculated from (\ref{res4}), accords well
with experimental data. The discrepancy between the theory and
experiment in the ferroelectric phase is obviously caused by the
contributions of the domain effects into the elastic constant of
the crystal, which are not considered in the present model. Note
that the first resonant frequency has a sharp minimum at the
transition point, owing to the similar behavior of  the elastic
constant $c_{66}^E$.

 \section{Resonant frequencies of NH$_{4}$H$_{2}$PO$_{4}$ type crystals}

We consider vibrations of a 0$^\circ$ Z-cut of an
antiferroelectric NH$_{4}$H$_{2}$PO$_{4}$ type crystal, produced
by an external time-dependent electric field $E_{3t}=E_3e^{i\omega
t}$. Taking into account the system Hamiltonian, the symmetry of
the proton distribution functions for the case of
antiferroelectric ordering \cite{old-adp,0819U}, and following the
procedure, described in the previous section, we obtain an
expression for the dynamic dielectric permittivity of a free
crystal, which is  formally the same as for the case of
ferroelectric ordering (\ref{dyn_susc}). However, the elastic
constant is  different
 \bea
 && c_{66}^E(\alpha\omega) = c_{66}^{E0} +  \frac{4\beta \psi_6}{v D} \Bigl[ - 2\psi_6 F^{(1)}(\alpha\omega) + \delta_{s6} F_s^{(1)}(\alpha\omega)
  + \delta_{16} F_1^{(1)}(\alpha\omega) - \delta_{a6} F_a^{(1)}(\alpha\omega) \Bigr] - \nonumber\\
 && - \frac{4\varphi_c^{\eta} f_6}{v D} \beta
 \Bigl[ - 2\psi_6 F^{(1)}(\alpha\omega) + \delta_{s6} F_s^{(1)}(\alpha\omega)  + \delta_{16} F_1^{(1)}(\alpha\omega) - \delta_{a6} F_a^{(1)}(\alpha\omega)
  \Bigr] + \nonumber\\
 && + \frac{4\beta \psi_6}{v D} f_6 - \frac{2\beta}{v D}
 \Bigl[ \delta_{s6}^2 a  + \delta_{16}^2 4b + \delta_{a6}^2 (1 + \cosh 2\tilde x)\Bigr]. \label{c66e-adp}
 \eea
Just like in the case of KH$_2$PO$_4$ type crystals, it does not
have any perceptible frequency variation in the piezoelectric
resonance region and coincides with the static constant
$c_{66}^E$. On the other hand, in NH$_4$H$_2$PO$_4$ the elastic
constant $c_{66}^E$ does not exhibit any anomalous behavior in the
transition region and is about $6\cdot 10^{11}$~N/m$^2$ between
$T_{\rm N}=148$~K and 300 K \cite{0819U}.

\begin{wrapfigure}{i}{0.55\textwidth}
\centerline{\includegraphics[width=0.5\textwidth]{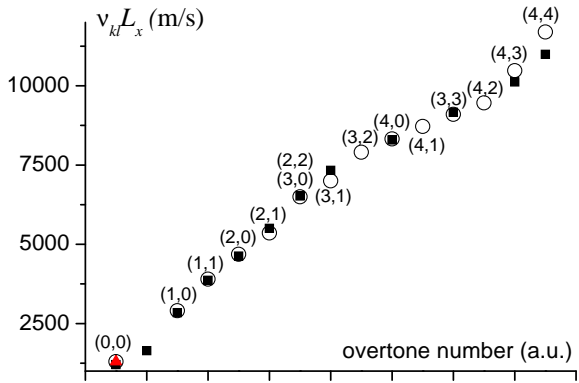}}
\caption{The frequency constants
$\nu_{kl}L_x=L_x\omega_{kl}^0/2\pi$ of a square 0$^\circ$ Z-cut of
a NH$_4$H$_2$PO$_4$ crystal. $\blacksquare$, $\blacktriangle$:
experimental points taken from \cite{Gainon} and \cite{Bechmann}.
$\circ$: the present theory. Numbers is parentheses are the
$(k,l)$ values. } \label{fig-adp}
\end{wrapfigure}

The expressions for the piezoelectric coefficient
$e_{36}(\alpha\omega)$,  dynamic dielectric susceptibility of a
clamped crystal $\chi_{33}^\eps$,  the function $ {R_6(\omega)}$,
and the equation for the resonant frequencies are the same as in
the case of a ferroelectric KH$_2$PO$_4$ type crystals:
(\ref{e36}), (\ref{chie}), (\ref{r6}), and (\ref{res4}),
respectively. However, the functions $F^{(1)}(\omega)$ and other
auxiliary quantities used in these formulae as well as in
(\ref{c66e-adp}) differ from those from the previous section and
can be found in \cite{old-adp,0819U}.

In fig.~\ref{fig-adp} we compare the calculated frequency
constants (the resonant frequencies multiplied by the sample edge
length $L_x\omega_{kl}^0$; the size-independent quantity) of a
square 0$^\circ$ Z-cut of a NH$_4$H$_2$PO$_4$ crystal to the
available experimental data. As one can see, a very good agreement
is obtained. The fitting procedure and values of the model
parameters were given in \cite{old-adp,0819U}.

\section{Conclusions}

Within the proton ordering model with taking into account the
shear strain $\varepsilon_6$ we explored a dynamic response of
ferroelectric and antiferroelectric crystals of the  KH$_2$PO$_4$
family to an external harmonic electric field $E_3$. Dynamics of
the pseudospin subsystem is described within the stochastic
Glauber approach. Dynamics of the strain $\varepsilon_6$ is
obtained from the Newtonian equations of motion of an elementary
volume, with taking into account the relations between the order
parameter of the pseudospin subsystem and the strain. Corrected
expressions for the piezoelectric resonance frequencies of simply
supported rectangular 0$^\circ$ Z-cuts of these crystals are
obtained. They are shown to yield a good quantitative agreement
with experimental data for KH$_2$PO$_4$ and NH$_4$H$_2$PO$_4$
crystals.

The ultimate goal of the present studies  will be to generalize
the obtained expression for the dynamic permittivity to the case
of the Rb$_{1-x}$(NH$_4)_x$PO$_4$ type  proton glasses, in order
to explore their dynamic dielectric response. It is known
\cite{Schmidt,De-Gao} that, just like their pure constituents,
these mixed systems are piezoelectric, and their dynamic
dielectric permittivity has a piezoelectric resonance dispersion.
As our preliminary calculations show, the experimentally obtained
resonant frequencies of such mixed crystals \cite{Trybula} are
well described by the obtained here expression for the resonant
frequencies, provided that the corresponding elastic constant
$c_{66}^E$ of such a system is known.

\section*{Acknowledgement}
The authors acknowledge support from the State Foundation for
Fundamental Studies of Ukraine, Projects ``The phase transitions
and physical properties of the KH$_2$PO$_4$-NH$_4$H$_2$PO$_4$
systems with competing ferroelectric and antiferroelectric
interactions'' No. F40.2/099 and ``Electromechanical nonlinearity
of mixed ferro-antiferroelectric crystals of dihydrogen phosphate
family'' No F53.2/070.


\end{document}